\documentclass[10pt, twoside, twocolumn]{IEEEtran}
\usepackage{atbegshi}
\usepackage[dvips]{graphicx,color}
\usepackage[cmex10]{amsmath}
\usepackage{array}
\usepackage[tight,footnotesize]{subfigure}
\usepackage{amsmath,amsfonts,graphicx,mathrsfs,subfigure,cite,url}
\usepackage{longtable,array,stfloats}
\usepackage[utf8]{inputenc}
\usepackage[keeplastbox]{flushend}
\usepackage{color}
\usepackage{multirow,multicol,makecell,booktabs}
\usepackage{tablefootnote}
\usepackage{verbatim}
\usepackage{arydshln}
\usepackage{tikz}
\usetikzlibrary{arrows,shapes,trees}
\usepackage{tikz-3dplot}

\newtheorem{proposition}{Proposition}

\newtheorem{theorem}{Theorem}

\makeatletter
\newcommand{\Rmnum}[1]{\expandafter\@slowromancap\romannumeral #1@}
\makeatother

\newcommand{\A}{\mathcal{A}}
\newcommand{\B}{\mathcal{B}}

\newcommand{\D}{\mathcal{D}}
\newcommand{\G}{\mathcal{G}}

\renewcommand{\H}{\mathcal{H}}
 \newcommand{\I}{\mathcal{I}}
\newcommand{\NS}{\mathcal{N}}

\newcommand{\R}{\mathcal{R}}

\renewcommand{\S}{\mathcal{S}}
\newcommand{\V}{\mathcal{V}}
\newcommand{\W}{\mathcal{W}}

\newcommand{\Z}{\mathbb{Z}}

\newcommand{\ha}{\widehat{a}}

\newcommand{\hs}{\widehat{s}}

\renewcommand{\d}{\mathrm{d}}

\newcommand{\sinc}{\mathrm{sinc}}

\newcommand{\con}{\mathrm{con}}

\newcommand{\opt}{\mathrm{opt}}

\newcommand{\reg}{\mathrm{reg}}
\newcommand{\sub}{\mathrm{sub}}
\newcommand{\lam}{\mathrm{\lambda}}

\newcommand{\nn}{\nonumber}

\definecolor{mycolor2}{rgb}{0,0.4,0.7}
\definecolor{mycolor}{rgb}{0,0.8,0}
\definecolor{mycolor1}{rgb}{.4,0.3,.13}

\begin{document}
\title{\Huge{{Constrained Sampling: Optimum Reconstruction in Subspace with Minimax Regret Constraint}}}

\author{Bashir Sadeghi , Runyi Yu,
and Vishnu Naresh Boddeti
\thanks{ \hspace{0cm}
1053-587X (c) 2019 IEEE. Personal use is permitted, but republication/redistribution requires IEEE permission.}

}

\markboth{\tiny{This article has been accepted for publication in IEEE Transactions on Signal Processing. Citation information: DOI 10.1109/TSP.2019.2925608, IEEE Transactions on Signal Processing.}}
{SADEGHI, YU, AND BODDETI: Constrained Sampling: Optimum Reconstruction in Subspace with Minimax Regret Constraint}

\maketitle
\begin{abstract}
This paper considers the problem of optimum reconstruction in generalized sampling-reconstruction processes (GSRPs).
We  propose constrained GSRP, a novel framework that minimizes the reconstruction error for inputs in a subspace,
subject to a constraint on the maximum regret-error for any other signal in the entire signal space. This framework addresses the primary limitation of existing GSRPs (consistent, subspace and minimax regret), namely, the assumption that the \emph{a priori} subspace is either fully known or fully ignored. We formulate constrained GSRP as a constrained optimization problem, the solution to which turns out to be  a convex combination of the subspace and the minimax regret samplings. Detailed theoretical analysis on the reconstruction error shows that constrained sampling achieves a reconstruction that is 1) (sub)optimal for signals in the input subspace, 2) robust for signals around the input subspace, and 3) reasonably bounded for any other signals with a simple choice of the constraint parameter. Experimental results on sampling-reconstruction of a Gaussian input and a speech signal demonstrate the effectiveness of the proposed scheme.
\end{abstract}

\begin{IEEEkeywords}
Consistent sampling, constrained optimization,
generalized sampling-reconstruction processes,
minimax regret sampling, oblique projection,
orthogonal projection, reconstruction error,
subspace sampling.
\end{IEEEkeywords}

\section{Introduction}

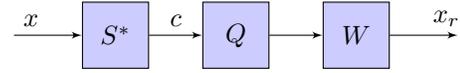
\begin{figure}
    \centering
        \tikzstyle{int}=[draw, fill=blue!20, minimum size=2.5em]
    \tikzstyle{init} = [pin edge={to-,thin,black}]
    \begin{tikzpicture}[node distance=1.6cm,auto,>=latex']
        \node [int] (a) {$S^*$};
        \node (b) [left of=a,node distance=1.35cm, coordinate] {};
        \node [int] (c) [right of=a] {$Q$};
        \node [int] (d) [right of=c] {$W$};
        \node [coordinate] (end) [right of=d, node distance=1.35cm]{};
        \path[->] (b) edge node [anchor=south east]{$x$} (a);
        \path[->] (a) edge node [anchor = south]{$c$} (c);
        \draw[->] (c) edge node {} (d);
        \draw[->] (d) edge node [anchor=south west]{$x_r$} (end);
    \end{tikzpicture}
    \caption{A typical GSRP: $S^*$ is a sampling operator, $Q$ is a {linear} discrete-time correction filter,  and $W$ a reconstruction operator.}
    \label{GSRP}
\end{figure}

Sampling is the backbone of many applications in digital communications and signal processing; for example, sampling rate conversion for software radio~\cite{Hentschel-2000}, biomedical imaging~\cite{Lehmann-1999}, image super resolution~\cite{Farsiu-2004}, machine learning and signal processing on graph~\cite{ortega-2018},~\cite{kovacevic-2015}, etc. Many of the systems involved in these applications can be modeled as the generalized sampling-reconstruction process (GSRP) as shown in Fig.~\ref{GSRP}. A typical GSRP consists of a sampling operator $S^*$ associated with a sampling subspace $\S$ in a Hilbert space $\H$, a reconstruction operator $W$ associated with a reconstruction subspace $\W\subseteq\H$, and a correction digital filter $Q$.
For a given subspace $\W$, orthogonal projection onto $\W$  minimizes the reconstruction error in $\W$, as measured by the norm of $\H$. As a result, orthogonal projection is considered to be the best possible GSRP. However, the orthogonal projection is not possible unless the reconstruction space is a subspace of sampling space~\cite{Eldar-2006}, i.e., $\W \subseteq \S$.
Therefore, many solutions have been developed for the GSRP problem under different assumptions on $\S$, $\W$ and the input subspace. These solutions can be categorized into \emph{consistent}, \emph{subspace}, and \emph{minimax regret} samplings.

When the inclusion property ($\W \subseteq \S$) does not hold, but  one still wants to have the effect of orthogonal projection for any signals in the reconstruction space, Unser \emph{et al}~\cite{Unser-1994,Unser-1998} introduced  the notion of \emph{consistent} sampling for shiftable spaces. This sampling strategy has later been developed and generalized by Eldar and co-authors~\cite{Eldar-2003-Jan,Eldar-2004,Dvorkind-2008,Eldar-Unser-2006}. Common to this body of work is the assumption that the subspace $\W$ and the orthogonal complement  $\S^\perp$ of subspace $\S$  satisfy the so-called direct-sum condition, i.e., $\W \oplus \S^\perp =\H$. This implies that $\W$  and $\S^\perp$ uniquely decompose $\H$. When the direct-sum condition is relaxed to be a simple sum condition $\W+\S^\perp = \H$, the consistent sampling can still be developed in finite spaces ~\cite{Hirabayashi-2007},~\cite{Dvorkind-2009}. Further generalization of consistent sampling where even the sum condition is not satisfied can be found in~\cite{Arias-2012,Kwon-2015}.

{In many instances and for various reasons, the reconstruction space $\W$ can be  different from the input subspace $\A$ which models input signals based on \emph{a priori} knowledge.
On  one hand, this may be the case due to limitation on physical devices.
On the other hand, it can also be advantageous to select suitable reconstruction spaces.
For example, band-limited signals are often used to model natural signals. In this case, the sinc function as a generator for the corresponding input space $\A$  suffers from slow convergence in reconstruction;
it is preferable to use a different generator that has short support (thus allowing fast reconstruction) for the reconstruction space $\W$.
}
Eldar and  Dvorkind in~\cite{Eldar-2006} introduced \emph{subspace} sampling and showed that orthogonal projection onto the reconstruction space
for signals belonging to \emph{a priori} subspace is feasible under  the direct-sum condition
between $\A$ and $\S^\perp$ (i.e., $\A \oplus \S^\perp =\H$).
The subspace $\A$ can be learned empirically or by a training dataset~\cite{Knyazev-2017}.
Nevertheless, it would still be subject to uncertainties due to, for example, learning imperfection, noise or hardware inability to sample at Nyquist rate.
Knyazev \emph{et al} used
a convex combination of consistent and subspace GSRP to address the
uncertainty of the \emph{a priori } subspace~\cite{Knyazev-2017}.
However, the reconstruction errors of  consistent sampling and subspace sampling can be arbitrarily large
if the angle between reconstruction (or \emph{a priori}) subspace and sampling subspace approaches $90^\circ$~\cite{Eldar-2006}.

\emph{Minimax regret} sampling was introduced by Eldar and Dvorkind~\cite{Eldar-2006} to address the possibility of large errors associated with consistent (and subspace) sampling for signals away from the sampling subspace. It minimizes the maximum (worst)
regret-error (distance of the reconstructed signal from orthogonal projection).
The minimax regret sampling,  however, is found to be conservative
as it ignores the  \emph{a priori} information on input signals.

In the aforementioned GSRPs the \emph{a priori} subspace is assumed to be either fully known or fully ignored, which is not practically realizable. In addition,
the angle between sampling space and input space cannot be  controlled  (they can get arbitrarily close to $90^\circ$).
In this paper, we introduce \emph{constrained} sampling to address these limitations.
We design a robust (in the sense of angle between sampling and  input  spaces) reconstruction for the signals
that  approximately lies in  the \emph{a priori} subspace. To this end, we introduce a new sampling strategy that
exploits the \emph{a priori} subspace information while enjoying the reasonably bounded error (for any input) of  the minimax regret sampling.
This is done by minimizing the reconstruction error
for the signals lying in the \emph{a priori} subspace
while constraining the minimum regret-error to be below certain level for any signal in  $\H$.
The solution is shown to be  a convex combination of minimax regret and consistent sampling.
To be specific, given an input $x$, the reconstruction of the proposed constrained sampling
is given as a convex combination
\begin{equation}
x_\lam = \lam\, x_\sub+(1-\lam)x_\reg, \quad   \lam \in [0, \, 1]
\end{equation}
where $x_\sub$ and $x_\reg$ are the reconstructions of  the subspace and minimax sampling, respectively.
The result is illustrated in Fig.~\ref{fig-reconstructions-1}
for a  simple case where  $\H=\mathbb{R} ^2$ and the \emph{a priori} subspace $\A$ is equal to the reconstruction subspace $\W$
(therefore, subspace sampling is the same as consistent sampling).
In the figure, $x$ is the input signal; $x_\opt =P_\W x$ is the optimal reconstruction, i.e., the orthogonal projection of $x$ onto $\W$;
$x_\sub = P_{\W \S^\perp} x$
is the oblique projection onto $\W$ along the orthogonal complement of $\S$;
and $x_\reg = P_\W P_\S x$ is the result of two successive orthogonal projections.
The figure shows that as a combination of $x_\sub$ and $x_\reg$, our constrained sampling $x_\lam$
can potentially be very close to orthogonal projection.
This desirable feature will also be demonstrated in the two examples in Section~\ref{sec-F}.

\begin{figure}
  \centering
\tdplotsetmaincoords{0}{0}
\begin{tikzpicture}[scale=3.2,tdplot_main_coords]

\draw[thick,-] (0,0,0) -- (0,1.1,0) node[anchor=east]{$\S^\perp$};
\draw[thick,-] (0,0,0) -- (1.41,0,0) node[anchor=north]{$\S$};

\draw[thick,-] (0,0,0) -- (1.1,1.1,0) node[pos=0.97,above,sloped]{$\A=\W$};

\draw[thick,-,blue] (0,0,0) -- (1.25,0.625,0) node[anchor=west]{$\B$};
\filldraw[]  (0,0,0) -- (1,0.4,0)circle (.4pt) node[anchor=west]{$x$};
\filldraw[blue]  (1,0.5,0)circle (.4pt);
\filldraw[dashed,red]  (1,0.4,0) -- (1,0,0)circle (.4pt) node[anchor=north]{$P_\S x$};
\filldraw[dashed,red]  (1,0,0)-- (0.5,0.5,0) circle (.4pt) node[pos=.82,above,sloped]{$x_\reg$};
\filldraw[dashed,mycolor]  (1,0.4,0)-- (1,1,0) circle (.4pt) node[pos=.83,below,sloped]{$x_\sub$};
\filldraw[dashed,mycolor2]  (1,0.4,0)-- (.7,.7,0) circle (.4pt) node[pos=.7,below,sloped]{$x_\opt$};
\filldraw[dashed,blue]  (1,0.5,0)-- (.75,.75,0) circle (.4pt) node[pos=.7,above,sloped]{$x_\lam$};
\end{tikzpicture}
  \caption{Comparison of several sampling schemes: $x_\opt =P_\W x$ is the orthogonal projection of $x$ onto $\W$;
  $x_\sub=P_{\W \S^\bot}x$ is the oblique projection onto $\W$ along $\S^\perp$; and $x_\reg= P_\W P_\S x$ is the orthogonal projection onto $\S$ followed by orthogonal projection onto $\W$. Our constrained reconstruction $x_\lam$ is a simple convex combination of $x_\sub$ and $x_\reg$ and can be expressed as $ P_\W P_{B\S^\bot} x$ for some subspace  $\B$ given in Section~\ref{sec-C}.
  Note that $x_\lam$ can potentially get very close to $x_\opt$.}
\label{fig-reconstructions-1}
\end{figure}

The main contributions of this paper can be summarized as follows:
\begin{enumerate}
  \item We  propose and solve a constrained optimization problem which yields reconstruction that is
    (sub)optimal for signals in input subspace and robust for any other input signals.
  \item The solution to the  optimization problem leads to a new sampling strategy (i.e., the constrained sampling)
 which has  consistent (or subspace) and minimax regret samplings as  special cases.
  \item We  provide detailed analysis of reconstruction errors, and obtain reconstruction guarantees
  in the form of lower and upper bounds of errors.
\end{enumerate}

The organization of this paper is as follows.
In Sections~\ref{sec-B} and~\ref{sec-CC}, we provide preliminaries and discuss related work, respectively.
The proposed constrained sampling is described in Section~\ref{sec-C}.
In Section~\ref{sec-D}, we obtain lower and upper bounds on the reconstruction error of the constrained GSRP.
We then present two illustrative  examples to demonstrate the effectiveness of the new sampling scheme in Section~\ref{sec-F}.
Finally, we conclude the paper in Section~\ref{sec-G}.

\section{Preliminaries}
\label{sec-B}
\subsection{Notation}
We denote the set of real and integer numbers with $\mathbb R$ and $\mathbb Z$ respectively.
Let $\big(\H, \langle\,\cdot\,\rangle\big)$ be  a Hilbert space with the norm $\| \cdot \|$ induced by the inner product
$ \langle \, \cdot \, \rangle $.
We assume throughout the paper that $\H$ is infinite-dimensional unless otherwise stated.
Vectors in $\H$ are represented by lowercase letters (e.g., $x$, $v$).
Capital letters are used to represent operators (e.g., $S$, $W$).
The (closed) subspaces of $\H$ are denoted by capital calligraphic letters (e.g., $\S$, $\W$). $\S^\perp$ is the orthogonal complement of $\S$ in $\H$.
For  a linear operator $V$,  its range and nullspace are denoted by $\R(V)$ (or $\V$) and $\NS(V)$ respectively.
In particular, the Hilbert space of continuous-time square-integrable functions  (discrete-time summable sequences, resp) is denoted by $L_2$  ($\ell_2$, resp). At particular time instant $t\in \mathbb R$ ($n\in \Z$, resp), the value of signal $x\in L_2$ ($d\in \ell_2$, resp) is denoted by $x(t)$ ($d[n]$, resp).

\subsection{Subspaces and Projections}

Given two subspaces $\V_1,\, \V_2$, if they satisfy the direct-sum condition, i.e.,
\[
\V_1 \oplus \V_2 = \H
\]
we can define an oblique projection onto $\V_1$ along $\V_2$.
Let it be denoted as  $P_{\V_1 \V_2}$.
By definition~\cite{Eldar-2006}, $P_{\V_1 \V_2}$ is the unique operator satisfying
\[
 P_{\V_1 \V_2} \,x = \Bigl\{
       \begin{array}{ll}
                       x,  \quad  x\in\V_1 \\
                      0, \quad  x\in \V_2 .
       \end{array}  \Bigr.
\]
As a result, we have
\[
\R (P_{\V_1 \V_2})=  \V_1 , \quad  \NS (P_{\V_1 \V_2})=\V_2.
\]
Any projection $P$ can be written, in terms of its range and nullspace, as
\[
P=P_{\R(P) \NS(P)}.
\]

By exchanging the role of $\V_1$ and of $\V_2$,
we also have the oblique projection $P_{\V_2 \V_1}$. And
\begin{equation}\label{eq-x1}
  P_{\V_1 \V_2} + P_{\V_2 \V_1} = I
\end{equation}
where $I \colon \H \rightarrow \H$ is the identity operator.
In particular, if $\V_1=\V_2^\perp=\V$, then
the oblique projections reduce to the orthogonal ones, and (\ref{eq-x1}) specializes to
\begin{equation}\label{orthogonal-sum}
   P_{\V}  +  P_{\V^{\bot}} = I .
\end{equation}

An important characterization of projection is that a linear operator $ P\colon \H \rightarrow \H$
is an oblique projection if and only if $P^2 = P$~\cite{Christensen-2004}.
Note that the sum of two projections is generally not a projection.
Nevertheless, the following result states that their convex combination
remains  a projection if both share the same nullspace.
This result will be useful in our study of constrained sampling.

\begin{proposition}\label{prop-1}
Let $P_1$ and $P_2$ be two projections. If $ \NS(P_1) = \NS(P_2) $, then the following statements hold.
\begin{enumerate}
    \item $P_1 P_2 =P_1$ and  $P_2 P_1 =P_2$.
    \item  $P=\lam P_1+ (1-\lam) P_2$ is a projection for any $\lam\in \mathbb R$.
\end{enumerate}
\end{proposition}

\IEEEproof
 1) From (\ref{eq-x1}), it follows
  \[
        P_1 P_2 =  P_1 (I-P_{\NS(P_2) \R(P_2)}) = P_1 - P_1 P_{\NS(P_2) \R(P_2)}.
  \]
If $ \NS(P_1) = \NS(P_2) $, then the last term becomes zero.
Hence,  $P_1 P_2 =P_1$.
Similarly, we have that $P_2 P_1 =P_2$.

2) It can be readily verified that  $P^2=P$ in view of the result in~1).
\hfill \IEEEQEDopen

As  consequences of  Proposition~\ref{prop-1},   the  following equalities hold, which will be used in Section~\ref{sec-C}:
\begin{equation}\label{projection1}
  P_{\V_1} P_{\V_2 \V_1^{\bot}}  = P_{\V_1}
\end{equation}
and
\begin{equation}\label{projection2}
 P_{\V_1 \V_2^{\bot}} P_{\V_2}    =   P_{\V_1 \V_2^{\bot}}.
 \end{equation}

\subsection{Angle between Subspaces}

The notion of angles between two subspaces characterizes how far they are away from each other.

Consider a subspace $\V\subset \H$ and a vector $0 \neq x\in \H $.
The angle between $x$ and $\V$,
denoted by $(x, \V)$, is defined by
\begin{equation}\label{angle-vector}
\cos(x, \V) := \frac{\|P_\V x\|}{\|x\|}
\end{equation}
or equivalently
\begin{equation}\label{angle-sin1}
  \sin(x, \V) := \frac{\|P_{\V^{\bot}} x\|}{\|x\|}.
\end{equation}

Let $\V_1, \V_2 \subset \H$ be two subspaces,
following ~\cite{Eldar-2006}, the (maximal principal) angle between $\V_1$ and $\V_2$,
denoted by  $(\V_1, \V_2)$, is defined by
\begin{equation}\label{angle-cos}
\cos(\V_1, \V_2 ) :=  \inf_{0 \neq x \in \V_1} \frac{\|P_{\V_2} x\|} {\|x\|}
\end{equation}
or equivalently
\begin{equation}\label{angle-sin}
\sin( \V_1, \V_2 ) :=  \sup_{0 \neq x \in \V_1} \frac{ \| P_{\V_2^{\bot}} x\|}{\|x\|} .
\end{equation}
This angle can also be characterized via any linear operator $B$
whose range is equal to $\V_1$:
\begin{equation}\label{angle-c}
\cos(\V_1, \V_2 ) = \inf_{ x \not \in \NS(B)}  \frac{ \|P_{\V_2} B x\| }{\|B x\|}
\end{equation}
or equivalently
\begin{equation}\label{angle-s}
\sin( \V_1, \V_2 ) = \sup_{ x \not \in \NS(B)}  \frac{ \|P_{\V_2^{\bot}} B x\| }{\|B x\|}.
\end{equation}

Note that $  ( \V_1, \V_2 ) \neq ( \V_2, \V_1 )$ in general.
However, if their orthogonal complements are used instead, the order can be exchanged~\cite{Unser-1994, Eldar-2006}:
\begin{equation}\label{angle-commutative}
 ( \V_1, \V_2 )  =  (  \V^{\bot}_2, \V^{\bot}_1 ).
\end{equation}
Moreover,  under the direct-sum condition, commutativity holds~\cite{tang2000oblique}:
\begin{equation}\label{angle-communitative1}
  ( \V_1, \V_2 ) =( \V_2, \V_1 )  \quad \mbox{if } \V_1\oplus\V_2^\bot = \H  .
\end{equation}

The angle between subspaces allows descriptions of  lower and upper bounds
for orthogonal projection of signals in $\V_1$:
\begin{equation}\label{eq-x12}
    \cos(\V_1, \V_2)  \|    x \|    \le \|P_{\V_2}   x\|
    \le\sin(\V_1, \V_2^\bot)  \| x\|, \quad   x\in \V_1
\end{equation}
and for any signal in $\H$, via a linear operator $B$ with $\R(B)=\V_1$:
\begin{equation}\label{eq-x13}
    \cos(\V_1, \V_2)  \|   B  x \|    \le \|P_{\V_2} B x\|
    \le\sin(\V_1, \V_2^\bot)  \| B x\|, \;  x\in \H.
\end{equation}
For oblique projection, the following bounds are proven in~\cite{Eldar-2006}
\begin{equation}\label{consistent-bound}
\frac{\|P_{\V_2^\bot}x\|} {\sin(\V_1,\V_2)}
 \le \|P_{\V_1 \V_2}x\|
 \le \frac{\|P_{\V_2^\bot}x\|}{\cos(\V_1,\V_2^\bot)}.
\end{equation}
\section{Related Work}
\label{sec-CC}
In this Section, we review four important sampling schemes;
namely, orthogonal, consistent, subspace, and minimax regret samplings.
For comparison, some properties of these schemes, are summarized in Table~\ref{table-summary1},
along with the properties of our constrained sampling framework.

\setlength{\extrarowheight}{5pt}

\begin{table}[!t]
\begin{minipage}{0.50\textwidth}

\caption{Sampling schemes and their properties}
\label{table-summary1}
\begin{center}
\begin{tabular}{p{15mm} || p{19mm}|p{16mm}|p{15mm}}
\toprule
  \hfil Sampling& \hfil GSRP  & \hfil Optimal  & \hfil Error \\    [-1ex]
   \hfil Scheme & \hfil $T$  & \hfil in $\A$?   & \hfil bounded?\footnote{regardless of $(\A, \S)$.} \\
\hline
 Orthogonal\footnote{This is the optimal sampling scheme
 but possible only if $\W \subseteq \S$.}
 & \hfil $P_\W $  &  \hfil    optimal
  & \hfil bounded \\[1ex]
  \hdashline
 \hfil Consistent &\hfil $P_{\W \S^\bot}$  &  \hfil    optimal
  & \hfil unbounded \\
\hfil Subspace & \hfil $P_\W P_{\A \S^\bot}$ &  \hfil   optimal
  & \hfil  unbounded \\
 \hfil Regret & \hfil $P_\W P_\S$ &\hfil non-optimal
  & \hfil  bounded \\[1ex]
 Constrained \newline
  &  \hfil $\lambda P_\W P_{\A\S^\bot}$ \newline \hfil  $+(1-\lambda)P_\W P_\S$
  &  \hfil  sub-optimal
  &  \hfil bounded  \\[2ex]

\bottomrule
\end{tabular}
\end{center}

\end{minipage}
\end{table}

\subsection{Generalized Sampling-Reconstruction Processes}
\label{sec-Bd}

Consider the GSRP in Fig.~\ref{GSRP},
where  $x,\, x_r \in \H$ are the input and output signals, respectively;
$S^*$ and $W$ are the sampling and reconstruction operators, respectively;
and {$Q : \ell_2\rightarrow\ell_2$ is a bounded linear operator which acts as a correction filter}.

{Sampling and reconstruction spaces are usually restricted by acquisition and reconstruction devices or algorithms and are not free to be designed.}
Therefore, we assume that $S^*$ and $W$ are given in terms of sampling space $\S$ and reconstruction space $\W$, respectively.
Let $\W$  be spanned by a set  of vectors $\{ w_n \}_{n\in\I}$,
where $\I\subseteq \Z$ is a set of indexes.
Then $W \colon  \ell_2(\I) \rightarrow  \H$ can be described by the synthesis operator
\[
W \colon c \mapsto W c=\sum_{n\in \I} c[n] w_n, \quad c\in \ell_2 (\I).
\]
Note that  the range of $W$ is $\W$.

Similarly, let $\S$ be spanned by  vectors   $\{ s_n \}_{n\in\I}$.
Then $S^* \colon \H\rightarrow \ell_2(\I)$ can be described by the adjoint (analysis) operator
\begin{equation} \label{eq-sample}
S^* \colon x \mapsto   S^* x = c,  \quad c[n]=\langle x ,s_n\rangle, \quad n\in \I,  \, x\in \H
\end{equation}
since by definition of adjoint operator~\cite{kreyszig1978introductory}
\[
\langle S a, x\rangle = \langle a, S^* x\rangle_{\ell_2} \quad \text{for all } \ x\in \H,\, a\in \ell_2(\I).
\]
{In (\ref{eq-sample}), $c$ represents a sample sequence due to the sampling operation on $x\in \H$, i.e., $c=S^*x$. Note that if $ c = S^* x$ then for any input $v\in \S^\bot$ it holds $ c=S^* (x+v)$, since the orthogonal complement $\S^\bot$ is the  nullspace of $S^*$,
i.e.,  $\NS(S^*)= \S^{\bot}$~\cite{kreyszig1978introductory}.
}

We assume throughout the paper that set $\{w_n\}$ constitutes  a frame of $\W$, that is,
there exist two constant scalars  $0<\alpha \leq \beta <\infty$ such that
\[
\alpha \|x\|^2 \leq \sum_{n\in \I}| \langle x, w_n \rangle|^2\leq \beta \|x\|^2, \quad x\in\W.
\]
Set $\{s_n\}$ is also assumed to be a frame of $\S $.

The overall GSRP can be described as a linear operator $T \colon \H  \rightarrow \H$
\begin{equation}\label{eq-sysT}
  T     \colon   x \mapsto x_r = WQS^* x , \quad x\in \H.
\end{equation}

The reconstruction quality of the GSRP can be studied via the error system
 \begin{equation}\label{eq-sysE}
  E  := I- T = I-WQS^* .
\end{equation}
For any input $x\in \H$,
the reconstruction error signal is given as
\[
Ex= x - x_r .
\]

\subsection{Orthogonal Projection}
Consider the optimal reconstruction of signal $x$ by the GSRP in Fig.~\ref{GSRP}.
Since  $x_r \in \W$,
the (norm of)  error $ Ex $ is minimized by its orthogonal projection on $\W$:
\[
x_r= P_\W x
\]
and therefore, the optimal error system is
\begin{equation}
E_\opt  := I- P_\W = P_{\W^{\bot}}.
\end{equation}
For each $x\in\H$, the optimal error signal is
\begin{equation}
  E_\opt x = P_{\W^{\bot}} x.
\end{equation}

The orthogonal projection
$P_\W$ can be represented in terms of analysis and synthesis operators as~\cite{Eldar-2006}
\begin{equation}\label{projection-orthogonal}
P_{\W}=W (W^* W)^\dagger W^*
\end{equation}
where "$\dagger$" denotes the Moore-Penrose pseudoinverse.

According to~\cite{Eldar-2006},
$P_\W$ is subject to a fundamental limitation on the GSRP.
Specifically,
unless the reconstruction subspace is a subset of the sampling subspace, i.e.,
\begin{equation}\label{inclusion}
\W \subseteq \S
\end{equation}
there exists no correction filter $Q$ that renders the GSRP $T$ to be the orthogonal projection $P_\W$.

Acknowledging the optimality as well as the limitation of the orthogonal projection,
we now introduce the difference between $T$ and $P_\W$,
which is, in the spirit of~\cite{Eldar-2006},  referred to as the regret-error system:
 \begin{equation}\label{eq-Reg}
  R := P_\W- T  = P_\W - WQS^*.
\end{equation}
Then the regret-error signal is given as
\begin{equation}
R x =  P_\W x-x_r =  ( P_\W -W Q S^* )x.
\end{equation}

It is important to note  that the two error systems are related as
\begin{equation}\label{eq-s1}
E =R + P_{\W^\bot} .
\end{equation}

As the optimal sampling, orthogonal projection $P_\W$ enjoys the following two desirable properties:
\begin{enumerate}
  \item Error-free in $\W$:   i.e., $E x =0$ for any $x\in \W$; and
  \item Least-error for  $x\in \H$: i.e.,  $E x= E_\opt x$   for any   $x \in H$.
\end{enumerate}
Consequently, $\|Ex\| \leq \|x\|$ for any $x \in \H$.

\subsection{Consistent Sampling}
Consistent sampling achieves  the property of being error-free in $\W$
without requiring the inclusion condition (\ref{inclusion}) for the orthogonal projection.

Under the assumption of the following direct-sum condition
\begin{equation}\label{direct-sum}
\W \oplus  \S^{\bot} = \H.
\end{equation}
it is shown~\cite{Eldar-2003-Jan} that the correction filter
\begin{equation}\label{correction-oblique}
Q_{\con}: =(S^*W)^\dag
\end{equation}
leads to an error-free reconstruction for input signals in $\W$.

The resulted  GSRP is found to be  an oblique projection
\begin{equation}\label{projection-oblique}
T_\con :=  W(S^*W)^\dag S^* =  P_{\W {\S^\bot}}.
\end{equation}
As a result, it is sample consistent, i.e.,
\[
S^*(T_\con x)=S^*(x-P_{\S^\perp\W})=S^*x, \ x\in \H
\]
where we used~(\ref{eq-x1}) and the fact that $\NS(S^*)=\S^\perp$.

The consistent error system is
\begin{equation}\label{eq-x11}
E_\con := I- P_{\W  \S^\bot } =  P_{ \S^\bot \W}
\end{equation}
and the corresponding regret-error system  also has a simple form:
\begin{equation}\label{eq-c2}
  R_\con   := P_\W P_{\S^{\bot} \W}
\end{equation}
since, from (\ref{eq-Reg}), (\ref{orthogonal-sum}),
and~(\ref{projection2}), we have
\begin{eqnarray*}\label{error-consistent}
R_\con & = & P_\W  -  T_\con   \\
         & =  &   P_\W -  P_{\W S^\bot}  \\
         & =  &  P_\W  - P_\W P_{\W S^\bot} \\
         & =  & P_\W (I - P_{\W \S^{\bot}}) \\
         & =  & P_\W   P_{\S^{\bot} \W} .
\end{eqnarray*}
Therefore, $ E_\con x =  R_\con  x =0$ for any $x\in\W$.

The absolute error for any input can be derived as follows:
\begin{equation}
  \|E_\con x\|^2 = \|P_{\W^\bot}\,x\|^2 +\|P_\W P_{{\S^\bot}\W}\,x\|^2, \quad x \in \H.
\end{equation}
And the regret-error is
\begin{equation}
\| R_\con x\| = \|P_\W P_{{\S^\bot}\W}\,x\|, \quad x \in \H.
\end{equation}
From~\cite{Eldar-2006}, the absolute error can be bounded in terms of the subspace  angles as
\begin{equation}\label{error-consistent-bound2}
\frac{E_\opt x }{\sin(\W^\bot,\S)} \leq \| E_\con x \| \leq \frac{E_\opt x}{\cos(\W,\S)}.
\end{equation}
The regret-error is shown in Section~\ref{sec-C} to be bounded as
\begin{equation}\label{error-consistent-bound}
         \frac{\cos  (\W^\bot, \S )} {\sin  ( \W^{\bot},\S )} \|P_{\W^{\bot}}x\|
\leq   \|   R_\con x  \| \leq
          \frac{\sin ( \W, \S) }{\cos  (\W,\S) }  \|P_{\W^{\bot}}x\| .
\end{equation}

It is clear from the left-hand sides of (\ref{error-consistent-bound2}) and (\ref{error-consistent-bound})
that the absolute error and regret-error for $x\in \W^\perp$ can be arbitrarily large if angle $(\W^\bot,\S)$ approaches  to zero.

\subsection{Subspace Sampling}
The result on consistent sampling in the preceding section
{has been extended in~\cite{Eldar-2006}} to any input subspace $\A\subset \H$
that satisfies the direct-sum condition with $\S^\bot$, i.e., $\A \oplus \S^{\bot} = \H$.

{Recall that subspace $\A$ models the input signals based on our \emph{a priori} knowledge.
}
Let $\{a_n\}$ be a frame of subspace $\A$. Denote the corresponding  synthesis operator by $A$.
Then the correction filter
\begin{equation}\label{correction-subspace}
Q_{\sub} := (W^*W)^\dag W^*A(S^*A)^\dag.
\end{equation}
renders the GSRP to be the product of two projection operators:
\begin{equation}\label{eq-c5}
T_\sub : =   W  (W^*W)^\dag W^*A(S^*A)^\dag S^* =  P_\W   P_{\A \S^{\bot}}.
\end{equation}

The regret-error system now is
\begin{equation}\label{eq-c4}
  R_\sub  :=    P_\W - T_\sub = P_\W - P_\W P_{\A \S^\bot} = P_\W  P_{ \S^\bot \A} .
\end{equation}
And the error system is
\begin{eqnarray}\label{error-sub}
E_\sub := P_{\W^\bot} +P_\W  P_{ \S^\bot \A}.
\end{eqnarray}

Accordingly, the absolute error  and the regret-error are given, respectively,  by
\[
\| E_\sub x\|^2 = \|P_{\W^\bot}  x\|^2 +\|P_\W P_{{\S^\bot} \A}\,x\|^2, \quad x \in \H
\]
and
\[
\| R_\sub x\| = \|P_\W P_{ \S^\bot \A}\,x \|, \quad x \in \H.
\]
And the regret-error verifies the following error bounds:
 \begin{equation}\label{error-sub-bound}
         \frac{\cos  (\W^\bot, \S )} {\sin  (\A^\bot,\S )} \|P_A^\bot x\|
\leq   \|   R_\sub x  \| \leq
          \frac{\sin ( \W, \S) }{\cos  (\A,\S) }  \|P_A^\bot x\|
\end{equation}
which will be shown in Section~\ref{sec-D}.

For any $x\in \A$, it holds $P_{\S^{\bot} \A} x= 0$, thus $E_\sub x = E_\opt x$ and $R_\sub x=0$.
This implies that the optimum reconstruction is achieved for any $x\in\A$.
However,  the reconstruction error of $E_\sub x$ for $x \in \A^\perp$ can still be excessively large if angle $ (\A^\bot,\S )$ is very small,
which can be seen from (\ref{error-sub-bound}).

Recall that filter $Q_\sub$ is the minimizer of the reconstruction error for any input $x \in \A$; it is the solution to the following optimization problem~\cite{Eldar-2006}:
{\begin{equation}
 \label{optimization-subspace}
 \min_{Q} \| E x\|, \quad x\in \A .
 \end{equation}
}

\subsection{Minimax Regret Sampling}
Introduced in~\cite{Eldar-2006}, the minimax regret sampling
alleviates the drawback of large error associated with the consistent and subspace samplings.
This is achieved by  minimizing the maximum regret-error rather than the absolute error.

Consider the optimization problem:
\begin{equation}\label{minimax-original}
\min_{Q} \,\max_{x\in \mathcal{D}}\| R x\|
\end{equation}
where
\begin{equation}\label{sample-set}
\mathcal{D}:=\{x\in \H \colon \|x\|< L, c=S^*x\}
\end{equation}
{where scalar $L>0$ is introduced as a norm bound to limit contribution of inputs $x\in \S^\bot$ to  ensure that the maximum regret error in~(\ref{minimax-original}) is bounded, and $L$ should also be  sufficiently large to render $\D$ non-empty.
Interestingly, the solution to~(\ref{minimax-original}) is shown to be independent of $L$~\cite{Eldar-2006}.}
And the minimax regret solution is found to be
\begin{equation}\label{correction-regret}
Q_\reg := (W^*W)^\dag W^* \, S(S^*S)^\dag .
\end{equation}
Consequently,  the GSRP becomes the product of two  orthogonal  projections:
\begin{equation}\label{sys-reg}
T_\reg := W Q_\reg  S^*  = P_\W P_{\S } .
\end{equation}
Hence, the regret-error system is
\begin{equation}\label{eq-d7}
  R_\reg :=  P_\W  - T_\reg = P_W P_{\S^\bot}.
\end{equation}
And the error system is
\begin{equation} \label{error-regret}
E_\reg  := P_{\W^\bot} + P_\W  P_{\S^\bot}.
\end{equation}
Moreover,  the  regret-error is shown~\cite{Eldar-2006} to be bounded as
\begin{equation}\label{error-regret-bound}
  \cos  ( \W^\bot, \S  ) \|P_{\S^\bot} x\|
  \leq    \| R_\reg x\|
 \leq   \sin   (\W,\S  ) \|P_{\S^\bot }x \|.
\end{equation}
Clearly,
\begin{equation}
    \| R_\reg x\| \leq   \|x\|  ,  \quad  x\in \H.
\end{equation}
And
\begin{equation}
   \|E_\reg x \| \leq  \sqrt{2} \|x\|, \quad x\in \H
\end{equation}
since
\[
\|E_\reg x\|^2 \le \big({1+\sin^2(\W,\S)}\big) \|P_{\S^\bot }x \|^2.
\]
The above error estimates imply that $T_\reg$ results in good reconstruction for $x\in \H$,
at the cost of introducing error for $x\in \W$ (or $\A$).
Since  $T_\reg$ does not differentiate any input signals, it could be very conservative for signals in the input subspace.

\section{Constrained Reconstruction}
\label{sec-C}

Suppose that we know  \emph{a prior} that  input signal $x$
is close to $\A$ (i.e., $(x,A)$ is  small), but not necessarily  lies in $\A$.
This is relevant since in many practical scenarios,  input signals cannot be exactly modeled as elements in $\A$.
For example when $\A$ is learned via training set and only approximately described as an input subspace.
It is also technically necessary when,
for example, the sampling hardware is unable to sample at Nyquist rate or the input signal is only approximately bandlimited.
We can seek a correction filter to improve the conservativeness of the regret sampling,
and in the meantime to achieving minimum error for each $x\in \A$ as in the case of subspace sampling.
In other words, we wish to reach a trade-off between achieving the  two properties of orthogonal projection $P_\W$.
{It should be noted that we assume that the direct sum property (i.e., $\A \oplus\S^\perp =\H$)  holds throughout the paper.}

For this end, we propose the following optimization problem
\begin{gather}\label{objective-first-subspace}
\min_{Q} \| Ex  \|  , \quad x \in \A \cap \D \\
\mathrm{s.t.} \ \max_{x\in\D} \|R x\| \leq \beta_0 (c) \nn
\end{gather}
where $\D$ is given in~(\ref{sample-set}),
and $\beta_0$ represents an appropriate bound that is dependent on the sampling sequence $c$.
By restricting that $x$ belongs to $\D$, we imply that all such input signals give the same sequence $c$ which is assumed to be given (see~\cite{Eldar-2006}). Our problem is to find a correction filter $Q$ that minimizes the reconstruction error subject to the minimax regret constraint.
We note that the union of such $\D$'s for all $c\in \mathcal{R}(S^*)$ is equal to the entire signal space $\H$.
The above optimization problem (\ref{objective-first-subspace}) encapsulates two desiderata, (1) optimum reconstruction in $\A$ through the objective, and (2) minimax recovery for all inputs in $\H$ through the constraint.

The regret-error in the above constraint
can be relaxed with the error between the GSRP itself and the minimax regret reconstruction (rather than the orthogonal projection), i.e.,
\begin{equation}
{\max_{x\in \D} \|P_\W P_\S x-WQS^*x \| = \|P_\W S(S^*S)^\dag c -WQc \|.}
\end{equation}
Not only would this realization allow a simple and elegant solution to our search for an alternative sampling scheme,
it is also supported by the following arguments.
On one hand, from triangular inequality, we have
\begin{eqnarray}
\max_{x\in\D}\| R x\| & =  & \max_{x\in\D}\|P_\W x-WQS^*x\| \nn \\
&\leq &  \max_{x\in\D}\bigl\{\|P_\W P_\S x-WQS^*x \|
\nn\\
&& \hspace{7ex}+ \|P_\W x-P_\W P_\S x\|\bigr\} \nn\\
& = &\|P_\W S(S^*S)^\dag c-WQc\|\nn \\
&&  + \max_{x\in\D}\|P_\W x-P_\W P_\S x\| \label{inequaliy-traingular} .
\end{eqnarray}
On the other hand, it is shown in Appendix~\ref{app-a} that
\begin{eqnarray}\label{eq-t1}
\max_{x\in\mathcal{D}}\|R x\| &    \ge  &
 \frac1{\sqrt{2}} \bigl( \|P_\W S(S^*S)^\dag c-  WQc\|  \nn  \\
& & +\max_{x\in\D} \|P_\W x-P_\W P_\S x\|   \bigr) .
\end{eqnarray}
We complete the argument by noting that the last terms in  (\ref{inequaliy-traingular}) and (\ref{eq-t1})
are  independent of correction filter $Q$.

In view of the above discussions, we now  present the constrained optimization problem as follows:
\begin{gather}\label{objective-second-subspace}
{\min_{Q} \|x - WQc\| ,\quad x\in \A  \cap\D }\\
\mathrm{s.t.} \ \|P_\W S(S^*S)^\dag c-WQc\|  \leq \beta_1 (c). \nn
 \end{gather}
which would lead to an adequate approximation of the optimization problem in (\ref{objective-first-subspace}).
The upper bound $\beta_1(c)$ in (\ref{objective-second-subspace}) needs to be properly chosen.
Let us consider two extreme  cases: $\beta_1(c)=0$  and $\beta_1(c)=\infty$.
If $\beta_1(c)=0$, the strict constraint implies that the solution to~(\ref{objective-second-subspace})
is the standard minimax regret filter in (\ref{correction-regret}).
On the other hand, if  $\beta_1(c)=\infty$ (i.e., the constraint is removed),
then the objective function in
(\ref{objective-second-subspace}) is minimized by the correction filter $Q_\sub$ of the subspace sampling, which is  given in~(\ref{correction-subspace}).
Hence,
the upper bound of the constraint  in~(\ref{objective-second-subspace}) becomes
\begin{equation}\label{upper-bound}
\beta (c)  = \|P_\W S(S^*S)^\dag c-P_\W A(S^*A)^\dag c\| .
\end{equation}

From the above discussions,
we conclude that the upper bound in~(\ref{objective-second-subspace}) can be set to be
$ \beta_1(c)= \lam \beta (c)$ for some parameter $ \lam \in [0, 1]$.
Accordingly, we present the constrained optimization problem~(\ref{objective-second-subspace}) and its solution in the next theorem.

{
\begin{theorem}\label{thm-con}
Consider the constrained sampling problem
\begin{gather}\label{objective-final-subspace}
\min_{Q} \|x - WQc\| ,\quad x\in \A \cap \D \\
\mathrm{s.t.} \ \|P_\W S(S^*S)^\dag c-WQc\| \leq \lam \beta (c)  . \nn
\end{gather}
A solution to it is given as
\begin{equation}\label{correction-regularized}
Q_\lam := \lam Q_\sub + (1-\lam)Q_\reg .
\end{equation}
\end{theorem}

\hspace{3ex} \emph{Proof}:
It is proved in Appendix~\ref{app-b}. \hfill \IEEEQEDopen
}

Following Theorem~\ref{thm-con}, the constrained GSRP can be expressed as
\begin{equation}\label{oblique-new}
T_\lam := \lam T_\sub + (1-\lam) T_\reg .
\end{equation}

The constrained GSRP
$T_\lam$ can be simplified to have a simple expression. Define $B$ as the convex combination of two projections:
\begin{equation}\label{eq-t4}
B := \lam P_{\A \S^\bot } + (1-\lam) P_\S.
\end{equation}
In view of~(\ref{eq-c5}) and (\ref{sys-reg}), the GSRP can be further expressed compactly as
\begin{equation}
T_\lambda = P_\W  B.
\end{equation}

The next result
states that $B$ is in fact also an oblique projection with the nullspace
being $\S^\bot$.

\begin{proposition}\label{prop-x2}
The linear operator $B$ defined in~{(\ref{eq-t4})} is given as
\begin{equation}\label{eq-t2}
   B =  P_{\B \S^\bot } .
\end{equation}
where   $\B=\R(B) $.
\end{proposition}
\hspace{3ex} \emph{Proof}:
It is proved in Appendix~\ref{app-p2}.
\hfill \IEEEQEDopen

Following Proposition~\ref{prop-x2},
the resulting constrained GSRP can be nicely described  as  the  product of two projections:
\begin{equation}\label{GSRP-regularized}
T_\lam = P_\W  P_{\B S^\bot}.
\end{equation}
Then, the regret-error system is
\begin{equation}\label{eq-x7}
  R_\lam : =   P_\W P_{S^\bot \B}   .
\end{equation}
And the error system is given  as
\begin{equation}\label{error-lambda1}
 E_\lam : =   P_{\W^\bot }  +P_\W P_{\S^\bot \B}.
\end{equation}
In view of~(\ref{eq-s1}),
and similar to the case of subspace sampling,
the reconstruction error is given by
\begin{equation}\label{eq-x6}
\| E_\lam x\|^2 = \|P_{\W^\bot}  x\|^2 +\|P_\W P_{{\S^\bot} \B}\,x\|^2, \quad x\in \H
\end{equation}
and the regret-error is
\begin{equation}\label{eq-x8}
\| R_\lam x \| = \|P_\W P_{ \S^\bot \B}\,x \|,  \quad x\in \H .
\end{equation}

It is interesting to see that all the  GSRPs  discussed  have the same expression as in (\ref{GSRP-regularized}).
When $\lam =0$, then $\B=\S$ and $T_\lam =T_\reg$;
and when $\lam =1$, then $\B=\A$ and $T_\lam = T_\sub$,
which becomes $T_\con$ if additionally $\A=\W$.
This shows that our constrained sampling generalizes all the other three samplings.
{ Regarding these two particular values of $\lam$, we recall that
if the input signals can be precisely modelled by $\A$, then  the subspace sampling should be chosen for the reconstruction.
On the other hand, if no \emph{a priori} information about the input signal is available,
it is better to choose the minimax regret sampling.
}

The description of $T_\lam$ in (\ref{GSRP-regularized}) shows that the constrained sampling is essentially a subspace sampling with a new modified subspace $\B$, which is comprised of all the convex combinations of vectors of $\A$ and $\S$ according to (\ref{eq-t4}).
Thus $\B$ is closer to $\S$ than $\A$ is, i.e., $(\B, \S) < (\A, \S)$, leading to a more  robust sampling strategy (i.e., better reconstruction for signals not in $\A$; further explanations on this observation will be given in Section~\ref{sec-D} following the  error analysis).
A geometrical illustration of all the sampling schemes is provided in~Fig.~\ref{fig-reconstructions}.

\begin{figure}
  \centering
    \tdplotsetmaincoords{79}{147}
\begin{tikzpicture}[scale=2.2,tdplot_main_coords]
    \def\p{.8}
    \def\t{1.5}
                \filldraw[
        draw=red,%
        fill=red!20,%
    ]          (-.2*\t,-\t*1.65,0)
            -- (1.2*\t,-\t*1.65,0)
            -- (1.7*\t,\t/1.15,0)
            -- (-.2*\t,\t/1.15,0)
            -- cycle;
        \draw(2.1*\t/2,-2*\t/1.6) node{$\S$};

        \draw[thick] (0,0,0) -- (1*2.6,-1/3*2.6,1/4*2.6) node[anchor=south] {$\W$};
                \filldraw[
        draw=gray,%
        fill=gray!15,%
    ]
               (2*\p,-2.6*\p,2*\p)
            -- (3*\p,2.8*\p,3*\p)
            -- (0,1.6*\p,0)
            -- (0,-3.1*\p,0)
            -- cycle;
        \draw(1.65*\p,-2.3*\p,1.8*\p) node{$\A$};
        \draw[very thin,dashed,mycolor] (2,1,1) -- ( 2.0000,1,2.0000) node[mycolor,anchor=east] {$P_{\A\S^\bot}x$};
        \filldraw[mycolor] ( 2.0000,1,2.0000) circle (.4pt);
        \draw[thin,dashed,mycolor2] (2,1,1) -- ( 1.6331,-0.5444,0.4083)node [pos=.95,below] {$x_\opt$};
        \filldraw[black] (1.6331,-0.5444,0.4083) circle (.4pt);
        \draw[very thin,dashed] (0,0,0) -- (2,1,2.0000);
        \draw[thin,dashed,mycolor] (2.0000,1,2.0000)-- (2*0.9231,-2*0.3077,2*0.2308) node [pos=.9,above,sloped] {$x_{\sub}$};
        \filldraw[mycolor] (2*0.9231,-2*0.3077,2*0.23083) circle (.4pt);

        \draw[very thin,dashed,red] (2,1,1) -- (2,1,0) node[red,anchor=east] {$P_{\S}x$};
        \filldraw[red] ( 2,1,0) circle (.4pt);
        \draw[very thin,dashed] (0,0,0) -- (2,1,0);
        \draw[thin,dashed,red] (2,1,0) -- (2*0.7101,-2*0.2367,2*0.1775);
        \node [red,above right] (A) at (2*0.7101+.27,-2*0.2367+.3,2*0.1775){$x_{\reg}$};
        \filldraw[red] (2*0.7101,-2*0.2367,2*0.1775) circle (.4pt);
        \draw[thin,dashed,blue] (2,1,1.2)node[anchor=west]{$P_{\B\S^\bot}x$} -- (2*0.8379,-2*0.2793,2*0.2095)node [pos=.95,above,sloped] {$x_\lam$};
        \filldraw[blue] (2,1,1.2) circle (.4pt);
        \filldraw[black] (0,0,0) circle (.4pt) node[anchor=west]{$0$};
        \filldraw[blue] (2*0.8379,-2*0.2793,2*0.2095) circle (.4pt);
        \draw[thin] (0,0,0) -- (1*2,1/2*2,1/2*2) node[anchor=west]{$x$};
        \filldraw [black] (1*2,1/2*2,1/2*2)circle (.4pt);
        \draw[thick,dashed] (0,0,0) -- (0,0,.73);
         \draw[thick] (0,0,0.73) -- (0,0,1.75) node[anchor=south]{$\S^\bot$};
\end{tikzpicture}
  \caption{An illustration of sampling schemes: $\S$ is the sampling space, $\W$ is the reconstruction space and $\A$ is the input space.
  $x_\opt = P_\W x$, $x_\sub = P_\W P_{\A\S^\bot}x$,
  $x_\reg = P_\W P_\S x$,  and $x_\lam = P_\W {P_{\B\S^\bot}x}$
  where $P_{\B\S^\bot}=\lam P_{\A\S^\bot}+ (1-\lam)P_\S$.
  Note that the  constrained reconstruction  $x_\lambda$  has the  potential  to approach optimum reconstruction $x_\opt$. }
  \label{fig-reconstructions}
\end{figure}

It should be noted that since $P_{\S^\bot \B}$ is still an \emph{oblique} projection,
the error $Ex$ can still be very large in general. However, we shall show in the next section that this concern can be removed
by properly choosing the value of parameter $\lam$, one such choice is $\lam =\cos (\A, \S)$.

\section{Analysis on Reconstruction Errors}
\label{sec-D}

{
This Section presents error performance for the proposed constrained sampling. First, we compare  the reconstruction error of constrained sampling
with those of the subspace sampling and that of  minimax regret sampling.

\begin{proposition}\label{thm-1}
The reconstruction error of constrained sampling is upper-bounded by a convex combinations of the corresponding errors of the subspace and minimax regret samplings as follows:
\begin{equation}
    \|E_\lam x\| \leq \lam\|E_\sub x\|+(1-\lam)\|E_\reg x\|, \quad x\in \H .
\end{equation}
The regret error of constrained sampling is similarly upper-bounded:
\begin{equation}
    \|R_\lam x\| \leq \lam\|R_\sub x\|+(1-\lam)\|R_\reg x\| , \quad x\in \H.
\end{equation}
\end{proposition}

\hspace{3ex} \emph{Proof}:
In view of  definitions of the error systems involved, we have
\[
E_\lam =I-T_\lam  = \lam E_\sub  +(1-\lam) E_\reg
\]
and similarly
\[
R_\lam =P_\W -T_\lam  = \lam R_\sub  +(1-\lam) R_\reg  .
\]
The results then readily follow from the triangular inequality of norm.
\hfill \IEEEQEDopen

Proposition~\ref{thm-1} implies that the reconstruction error of constrained sampling can never be larger than the other two corresponding errors at the same time.

Next, we present bounds on the regret-error of constrained sampling $T_\lam$ by examining
regret-error system $R_\lam$.

\begin{theorem}\label{thm-2}
For any $x\in\H$, the regret-error of constrained sampling is bounded as
\begin{equation}\label{error-bound-final}
\alpha_\lam \|P_{\B^\bot }x\|  \leq \|  R_\lam   x\| \leq \beta_\lam  \|P_{\B^\bot }x\|
\end{equation}
where the scalars are
\[
\alpha_\lam = \biggl(1+\lam^2\frac{\cos^2\bigl(\A^\bot,\S)}{\sin^2\bigr(\A^\bot,\S)}\biggr)^{\!\frac{1}{2}}
 \cos  ( \W^\bot , \S )
\]
and
\[
\beta_\lam = \biggl(1+\lam^2\frac{\sin^2 (\A,\S)}{\cos^2 (\A,\S)}\biggr)^{\! \frac{1}{2}}\sin  (\W, \S ).
\]
\end{theorem}
}

\hspace{3ex} \emph{Proof}:
First of all, since $\R(P_{S^\bot \B})= \S^\bot$,
it follows from (\ref{eq-x8}) and (\ref{eq-x13}) that
\begin{equation}\label{error-second}
  \cos (\W^\bot , \S )\|P_{\S^\bot  \B } x\| \leq  \|R_\lam x\|
        \leq \sin\big(\W, \S\big)\|P_{\S^\bot  \B } x\|.
\end{equation}
Moreover,
from~(\ref{consistent-bound}) and (\ref{angle-commutative}), it follows that
\begin{equation}\label{eq-x20}
\frac{\|P_{\B^\bot }x\|}{\sin\big(\B^\bot ,\S\big)}
\leq \|P_{\S^\bot  \B } x\|
\leq \frac{\|P_{\B^\bot }x\|}{\cos\big(\B,\S\big)}.
\end{equation}
Consequently, the regret-error enjoys the following estimates
\begin{equation}\label{error-bound}
\frac{\cos ( \W^\bot , \S  )}{\sin (\B^\bot ,\S) }  \|P_{\B^\bot }x\|
\le   \| R_\lam x \|
 \le  \frac{\sin (\W, \S)  }{\cos (\B,\S) } \|P_{\B^\bot } x\|.
 \end{equation}

We complete the proof by  simplifying the above bounds using
the following estimates of the trigonometrical  functions involving subspace $\B$:
\begin{eqnarray}\label{inequality-cos}
\frac{1}{1+\lam^2\frac{\sin^2(\A,\S)}{\cos^2(\A,\S)}}
\leq \cos^2\big(\B,\S\big)
\leq \frac{1}{1+\lam^2\frac{\cos^2(\A, \S^\bot )}{\sin^2(\A, \S^\bot )}}
\end{eqnarray}
and
\begin{eqnarray}\label{inequality-sin}
\frac{1}{1+\lam^2\frac{\sin^2(\A ,\S )}{\cos^2(\A ,\S )}}
\leq \sin^2\big(\B^\bot ,\S \big)
\leq \frac{1}{1+\lam^2\frac{\cos^2(\A^\bot ,\S)}{\sin^2(\A^\perp,\S )}}
\end{eqnarray}
which are proved  in  Appendices~\ref{app-a1} and \ref{app-a2}, respectively.
\hfill \IEEEQEDopen

\setlength{\extrarowheight}{6pt}
\begin{table*}[!t]
\begin{minipage}{1.0\textwidth}

\caption{Sampling strategies and their regret-errors}
\label{table-summary}

\begin{center}
\begin{tabular}{p{14mm} || c|p{41mm}|p{19mm}|p{30mm}|p{30mm}}
\toprule
 Sampling   &  GSRP   &  \hfil Correction Filter
        &  \multicolumn{3}{ c } {Regret Error $ \| R x\|=\|P_\W x - T x\| $\footnote{The absolute error is given by
         $ \| Ex\|^2  = \|x-Tx \|^2  = \|P_{\W^\perp} x\|^2 + \|R x  \|^2 $.}} \\ [1ex]
        \cline{4-6}
   Scheme &    $T$  & \hfil  $Q$  &  \hfil Expression & \hfil Lower Bound & \hfil Upper Bound   \\    [1ex]
\hline
 Orthogonal\footnote{This is the optimal sampling scheme
 but possible only if $\W \subseteq \S$. The corresponding reconstruction error is $ \|Ex\|  =\|P_\W^\bot x\|$.}
 & $P_\W $  &  \hfil    $(W^*W)^\dagger W^* S(S^*S)^\dagger$
  &  \hfil  $ 0 $
   &  \hfil $0$
&  \hfil  $0$
\\  [2ex] \hdashline
 Consistent & $P_{\W \S^\bot}$  &  \hfil    $(S^*W)^\dagger $
  & $ \|P_\W P_{\S^\bot \W} x\| $
   &  \hfil   $ \frac{\cos (\W^\bot,\S)}{\sin (\W^\bot,\S)} \, \|P_{\W^\bot}x \|  $
&  \hfil   $\frac{\sin (\W,\S)}{\cos (\W,\S)}  \,  \|P_{\W^\bot}x \|  $  \\[2ex]

Subspace &  $P_\W P_{\A \S^\bot}$ &  \hfil   $(W^*W)^\dagger W^*A(S^*A)^\dagger$
  & \hfil  $\|P_\W P_{\S^\bot\A} x\| $
  & $ \hfil  \frac{\cos (\W^\bot,\S)}{\sin (\A^\bot,\S)} \, \|P_{\A^\bot}x\| $
 &  \hfil   $\frac{\sin (\W,\S)}{\cos (\A,\S)} \, \|P_{\A^\bot}x\| $ \\[2ex]

 Regret  &   $P_\W P_\S$  & \hfil  $(W^*W)^\dagger W^*S(S^*S)^\dagger$ &
 $\|P_\W P_{\S^\bot} x\| $    &     \hfil  ${\cos (\W^\bot,\S)} \, \|P_{\S^\bot}x\| $
 & \hfil ${\sin (\W,\S)}\, \|P_{\S^\bot}x\| $  \\[2ex]

 Constrained\footnote{The modified subspace is $\B =\R \bigl( \lam P_{\A \S^\bot } + (1-\lam) P_\S \bigr) $,  $\lam \in [0, 1]$. }
  &  $P_\W P_{\B\S^\bot}$
  &  \hfil  $\lam (W^*W)^\dagger W^*A(S^*A)^\dagger$ \newline
        $ + (1 -\lam)(W^*W)^\dagger W^*S(S^*S)^\dagger $
  &  $ \|P_\W P_{\S^\bot \B} x \|  $
  & \hfil
    $ \Bigl( 1+ \lam^2 \frac{\cos^2 (\A^\bot,\S)} {\sin^2 (\A^\bot,\S) } \Bigr)^{\!\frac{1}{2}}$
       \newline $ \times  \cos (\W^\bot,\S) \,  \|P_{\B^\bot}x\| $
  & \hfil   $ \Bigl( 1+\lam^2 \frac{\sin^2(\A,\S)}{\cos^2(\A,\S)}\Bigr)^{\!\frac{1}{2}} $
        \newline $   \times \sin (\W,\S)\,  \|P_{\B^\bot} x \| $  \\[2ex]
        \hdashline
 Constrained \newline \hfil $x\in \A$
  &  $ P_\W-(1-\lambda)P_\W P_{\S^\perp}$
  &  \hfil  $\lam (W^*W)^\dagger W^*A(S^*A)^\dagger$ \newline
        $ + (1 -\lam)(W^*W)^\dagger W^*S(S^*S)^\dagger $
  &  \hfil $(1-\lambda)$ \newline

  $ \times \|P_\W P_{\S^\bot} \,  x \|  $

  &
   $\hfil (1-\lambda)\cos (\A,\S^\perp)$
       \newline  \hfil  $  \times  \cos(\W^\bot,\S) \,  \| x\| $
  &   $ \hfil (1-\lambda)\sin(\A, \S)$
        \newline \hfil $ \times \sin (\W,\S)\,  \|  x \| $  \\[1ex]
\bottomrule
\end{tabular}
\end{center}

\end{minipage}
\end{table*}

Note that The  bounds in Theorem~\ref{thm-2} specialize those for the other sampling schemes if $\lam =0$ or $1$.
Furthermore, it is important to  point out that
$(\B,\S) \le (\A,\S)$ for any $\lambda \in [0,1] $,
since
{
\[
\cos^2\big(\B,\S\big) \geq
\frac{\cos^2(\A,\S)}{{\cos^2(\A,\S)}+\lam^2{ \sin^2(\A,\S)}}
\geq  \cos^2\big(\A,\S\big)
\]
in view of lower bound of~(\ref{inequality-cos}) and the inequality $\cos^2(\A,\S)+\lam^2 \sin^2(\A,\S) \leq 1$ for $\lambda \in [0,1] $.}
In other words,  the modified subspace $\B$ inclines towards
$\S$ than the input subspace $\A$ does.
This explains from another perspective why the
constrained sampling would generally lead smaller maximum possible error than subspace sampling.

It is pointed out that with a simple choice of parameter
\begin{equation}\label{lambda-sufficient}
0 \le \lam \leq {\cos(\A,\S)}
\end{equation}
the reconstruction error in~(\ref{error-bound-final}) is seen to be bounded as below:
\begin{equation}\label{eq-z5}
\|R_\lam x\| \le\sqrt{2}\|x\|, \quad x\in \H.
\end{equation}
Then, the absolute error is bounded as
\begin{equation}\label{eq-z6}
\|E_\lam x\|\le\sqrt{3}\|x\|, \quad x\in \H.
\end{equation}

Finally, we turn to bounds on reconstruction errors for signal in input subspace $\A$.
If $x\in \A$, then
\begin{eqnarray}
\|R_\lambda x \|  & = &\|P_\W P_{\S^\perp\B }  x\| \nn \\
& = & \|P_\W [\lambda P_{S^\perp \A}+(1-\lambda)P_{\S^\perp}]  x\|  \nn  \\
& = & (1-\lambda)\|P_\W P_{\S^\perp} x \| \nn \\
& \le &(1-\lambda) \sin(\S^\perp, \W^\perp)\|P_{\S^\perp}  x\| \nn
\end{eqnarray}
where the first step is from~(\ref{eq-x8}) and the second step is from~(\ref{eq-t4}).
Thus, using~(\ref{angle-commutative}) and~(\ref{eq-x12}), we obtain an upper bound on regret-error
\begin{equation}
   \|R_\lambda x\| \leq (1-\lambda) \sin(\A,\S)\sin(\W, \S)\| x\|, \; x\in \A.
\end{equation}
Similarly, we can also obtain a lower bound on regret-error:
\begin{equation}
 \|R_\lambda x\|   \ge (1-\lambda) \cos(\A,\S^\perp) \cos(\W^\perp, \S)
  \|  x\| ,\; x\in \A .
\end{equation}
It then follows, from~(\ref{eq-s1}),~(\ref{eq-x6}), and (\ref{eq-x12}), that the absolute error are bounded as
\begin{equation}
\alpha_\A \| x\|  \leq \| E_\lam x\| \leq \beta_\A  \|x\|, \quad x\in \A
\end{equation}
where the scalars are
\[
\alpha_\A = \bigl(\cos^2(\A, \W^\bot) + (1-\lambda)^2 \cos(\A,\S^\perp) \cos(\W^\perp, \S)  \bigr)^{\! \frac{1}{2}}
\]
and
\[
\beta_\A = \bigl(\sin^2(\A, \W) + (1-\lambda)^2\sin(\A,\S) \sin(\W, \S) \bigr)^{\! \frac{1}{2}} .
\]

Table~\ref{table-summary} summaries key results on all the sampling schemes considered in this paper.

\section{Examples}
\label{sec-F}

We now provide two illustrative examples which consider reconstruction of a typical Gaussian signal and a speech signal.
These examples demonstrate the effectiveness of the proposed constrained sampling.

\subsection{Gaussian Signal}
\label{sec-F-C}
{ Most natural signals are approximately band-limited and can be adequately modelled as Gaussian signals.}
We now consider reconstruction of a Gaussian signal of unit energy:
\begin{eqnarray}\label{gaussian-signal}
x = \Bigl(\frac{1}{{\pi \sigma}}\Bigr)^{1/4}\exp({\frac{-t^2}{2\sigma}}),
\end{eqnarray}
where $\sigma=0.09$.

Assume that sampling period $T$ is one (i.e., the Nyquist radian frequency is $\pi$) and the sampling space $\S$ is the
shiftable subspace generated by the B-spline of order zero:
\begin{eqnarray}\label{B-spline-0}
s(t)=\beta^0(t)=
\biggl\{ \begin{array}{ll}
   1,  &  t \in [-0.5, \, 0.5)\\
   0,   & \mbox{otherwise}.
\end{array} \Bigr.
\end{eqnarray}
In other words, $\S$ is spanned by frame vectors $\{\beta^0 (t-n)\}_{n\in\Z}$.
Since $x$ has its  $94\%$ of its energy in the content of frequencies up to $\pi$,
it is  reasonable to assume that $\A$ is the subspace of $\pi$-bandlimited signals.
In this situation, we have $\cos(\A,\S)=0.64$, which can be calculated by~\cite{Unser-1994}
\begin{equation}
\cos^2(\A,\S) = \!\!\!\!\! \!\! \inf_{\omega \in [0,2\pi)} \frac{\big|\sum_{n\in \Z}\hs\, ^*(\omega+2\pi n)\,\ha (\omega+2\pi n)\big|^2}
{{\sum_{n\in \Z}|\hs(\omega+2\pi n)|^2\,\sum_{n\in \Z}|\ha(\omega+2\pi n)|^2}}\nn
\end{equation}
where $``\, \widehat{\cdot}\, "$ represents the Fourier transform, and $ a(t)=\sinc(t)$.
We further assume that the reconstruction space $\W$ is the shiftable subspace generated by the cubic B-splines~\cite{unser-1999-Bspline}
\begin{eqnarray}
w(t)= \beta^3(t) = [\beta^0 \ast \beta^0 \ast \beta^0\ast \beta^0](t)
\end{eqnarray}
where $`` \ast" $ is the convolution operator.

\begin{figure}[!t]
\centering
\includegraphics[width =85mm,height=62mm]{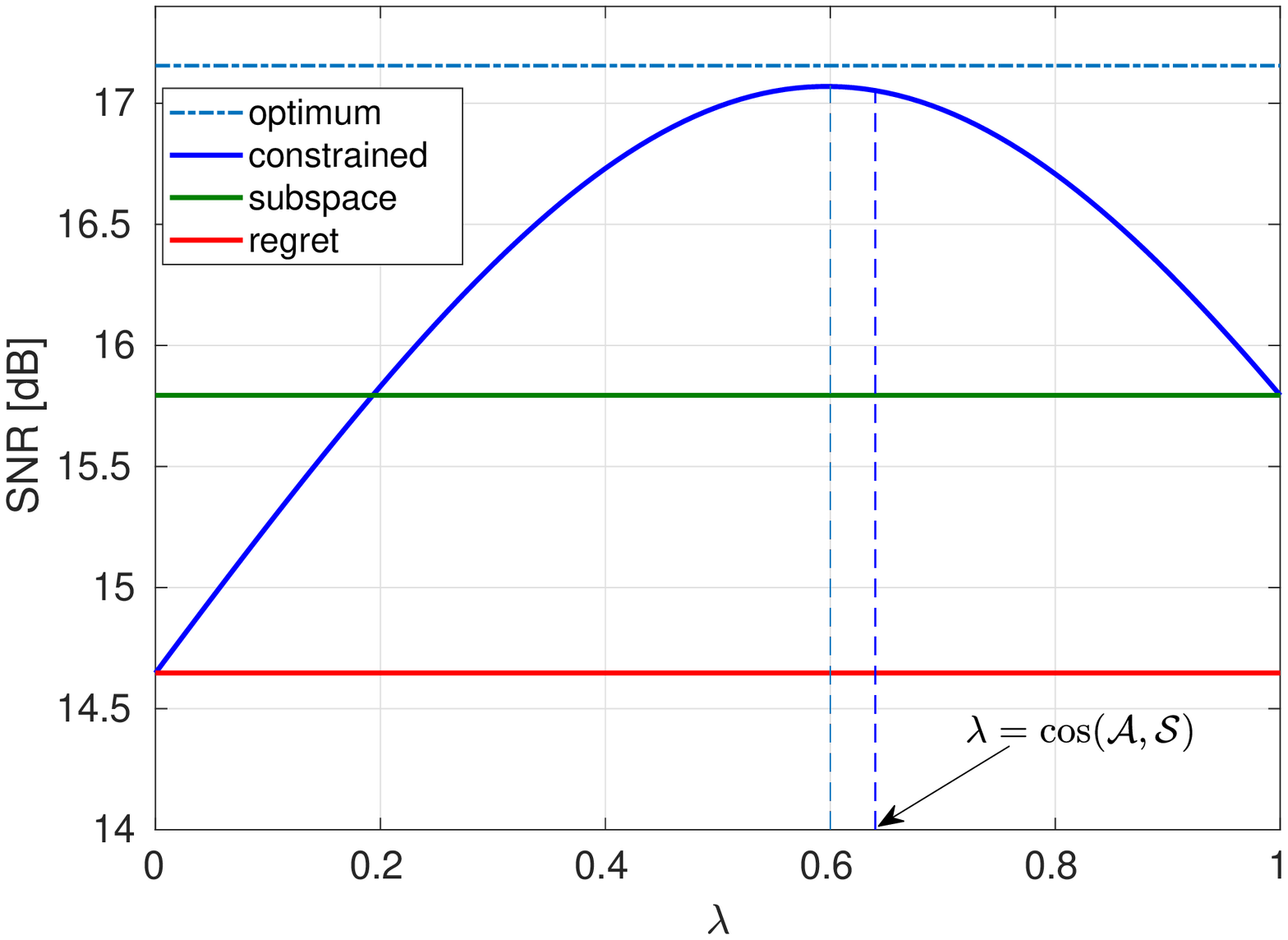}
\caption{Reconstruction error $\|Ex\|$ of a Gaussian signal for all four sampling schemes
($\S$ and $\W$ are generated by $\beta^0 $ and $\beta^3$, respectively,
 and $\A$ is the $\pi$-bandlimited subspace).}
\label{fig-exmp1}
\end{figure}

 Fig.~\ref{fig-exmp1} presents  the signal-to-noise ratio
 (SNR) in dB\footnote{ SNR$=20\log\big(\|x\|/\| Ex \| \big)$ dB}
 of the reconstruction error  $ Ex $  for the three sampling schemes.
We can observe from Fig.~\ref{fig-exmp1} that
1) the performance of the constrained sampling is never below that of the minimax regret
regardless of the value of $\lam$,
demonstrating the conservativeness of the regret sampling for inputs close to $\A$;
2) the constrained sampling achieves better reconstruction than
the subspace sampling for any $\lam  \in (0.20,  1)$;
3) with the simple choice of  $\lam=\cos(\A,\S)=0.64$,
the improvement of the constrained sampling over the subspace and minimax regret samplings are  $1.26$dB and $2.40$dB, respectively.

{
We recall that the Gaussian signal in (\ref{gaussian-signal}) is quite close to the $\pi$-bandlimited subspace $\A$ since $(x,A)=14.2^\circ$.
This closeness explains the worst performance of the minimax regret sampling which does not take advantage of any \emph{a priori} information on input $x$. The SNR of minimax sampling is less than the SNR of subspace by $1.14$dB.
On the other hand, since $x$ does not completely belong to $\A$, the performance of subspace sampling has also been improved by our constrained sampling which is capable of limiting the reconstruction error due to the content of frequencies beyond $\pi$. The improvement can be significant if parameter $\lam$ is properly selected.}
Furthermore,
it is  worth pointing out the existence of the optimal value (i.e., $\lam =0.60 \approx \cos(\A, \S)$)
such that  $\|E_\lam x\|$ is very close to (less than by $0.08$dB) the optimal error  $\|E_\opt x\|$,
demonstrating high potential of constrained sampling in approaching the orthogonal projection.

\subsection{Speech Signal}

In this example, the input signal is chosen to be a speech signal\footnote{downloaded from https://catalog.ldc.upenn.edu/}
which is sampled at the rate of $16$kHz.
Since the sampling rate is sufficiently high, the discrete-time speech signal $x[n]$ can accurately approximate
the continuous-time signal $x(t)$ on the  fine grid.
We assume that the sampling process $\S^*$ is an integration over one sampling duration $T$:
\begin{equation*}
c[n] = \frac{1}{T}\int_{nT-T/2}^{nT+T/2} x(t) \d t,
\end{equation*}
where $T=4000^{-1}$sec.
This is equivalent to assuming $s(t)=(1/T)\beta^0(t/T)$ or discrete-time filtering on the fine grid with filter whose impulse response is
\begin{equation*}
s[k] =
        \biggl\{ \begin{array}{ll}
           \frac{1}{3},  & k=-1,\,0,\,1 \\
            0,  & \text{otherwise}.
        \end{array} \Bigr.
\end{equation*}
Since the original continuous-time signal is sampled at $16$kHz,
we assume that subspace $\A$ is the space of $8$kHz-bandlimited signals.
For calculation,  we use a zero-phase discrete-time FIR low pass filter with
cutoff frequency at $1/2$ and of order $100$ to simulate $\A$ on the fine grid.
The selected $\A$ is equivalent to continuous-time low-pass filter with support $t\in$ $[-25T,\ 25T]$ which approximates $\sinc(4t/T)$.
For the synthesis, we let $w_n (t) =w(t-nT)$, where  $w(t)$ is chosen to have a time-support of
$t\in [-4T,\,4T]$ and to render a low pass filter with cutoff frequency (i.e., Nyquist frequency)
$1/ (2T) $.
On the fine grid, this synthesis process is  implemented via a discrete-time
low-pass FIR filter of order $16$ and with cutoff frequency $1/8$.

\begin{figure}[!t]
\centering
\includegraphics[width =85mm,height=62mm]{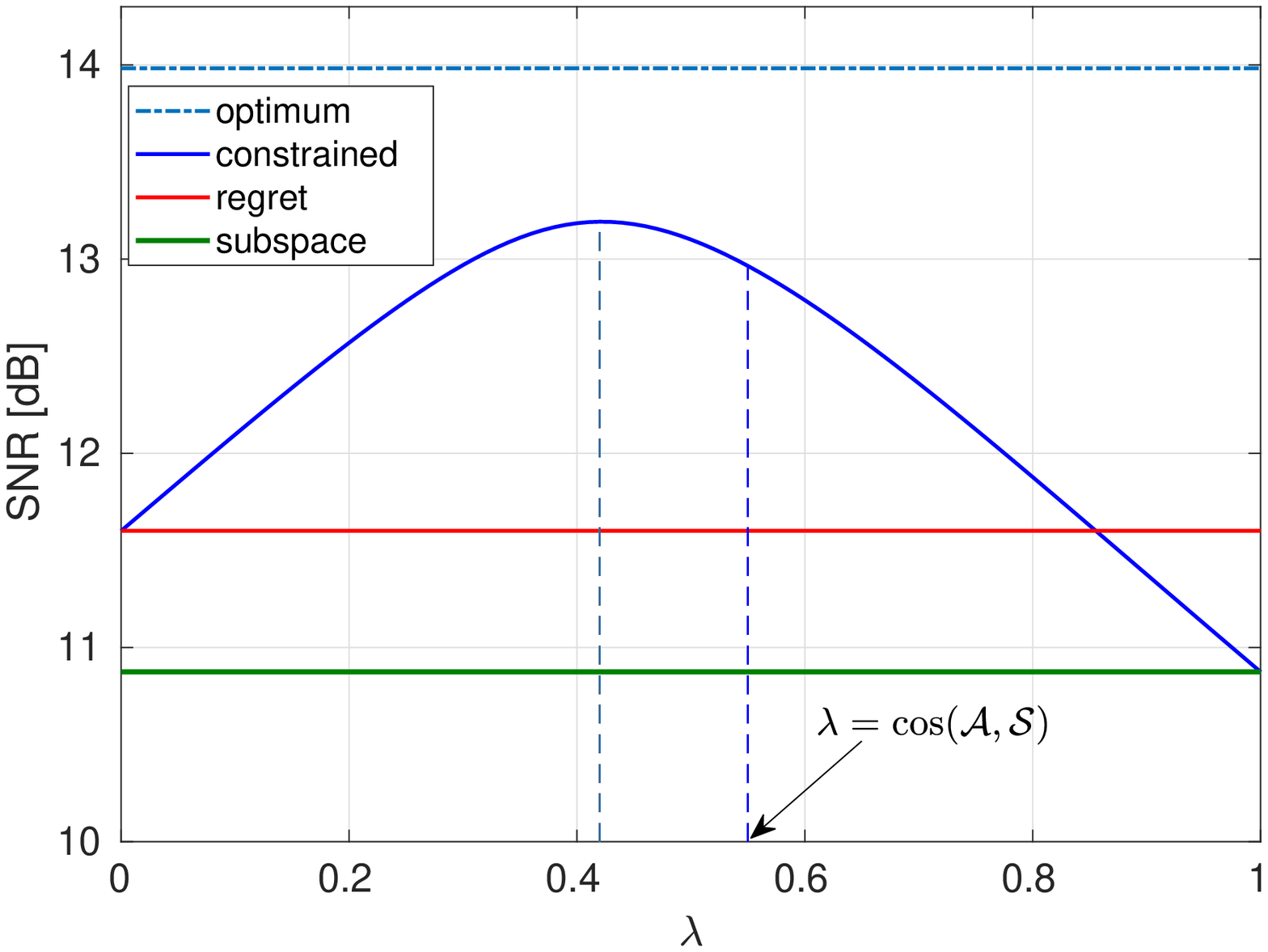}
\caption{Reconstruction error $\|E x\|$ of a speech signal for all four sampling schemes
($\S$ is generated by $\beta^0(t/T)$, $\W$ is generated by a non-ideal low-pass filter associated with a time-support of $[-4T, 4T]$, $T=4000^{-1}$s,
and $\A$ is the $8$kHz-bandlimited subspace).}
\label{fig-exmp2}
\end{figure}

In the experiment, following~\cite{Eldar-2006}, we randomly chose $5000$ segments (each with  $400$ consecutive samples) of the speech signal.
The segments are found to be far away from the \emph{a priori} $\A$ since the angles between them and $\A$ are found to be around $47.9^\circ$. Fig.~\ref{fig-exmp2} shows the reconstruction errors (averaged over all the segments) of the three sampling schemes.
As expected, the minimax regret sampling outperforms the subspace sampling (by $0.73$dB); and accordingly our constrained sampling always outperforms the subspace sampling (see also Proposition~\ref{thm-1}).
Moreover, when $\lam  \in [0, 0.85]$, the constrained sampling also outperforms the minimax regret sampling. For example, with a simple choice of  $\lam=\cos(\A,S)=0.55$, the improvement over the minimax regret and subspace samplings are  $1.37$dB and $2.10$dB, respectively.
Also note that at the optimum value of $\lam = 0.42$, the reconstruction error of
the constrained sampling is only $0.78$dB away from that of the orthogonal projection.
This result again shows the potential of the constrained sampling in approaching the optimal reconstruction.

{
The two examples above clearly demonstrate the effectiveness of the proposed constrained sampling over the minimax regret and subspace samplings when all input signals can be modelled (properly to some extent  but not precisely) by a subspace.
The results show that the constrained sampling is robust to model uncertainties and that it can potentially approach the optimal reconstruction when  parameter $\lambda$ is made adaptive to input characteristics even if the input is away from the input subspace.
}

\section{Conclusions}
\label{sec-G}
This paper re-examined the sampling schemes for  generalized sampling-reconstruction processes (GSRPs). Existing GSRP, namely, consistent, subspace, and minimax regret GSRPs, either assume that the input subspace is fully known or it is completely ignored. To address this limitation, we proposed, \emph{constrained sampling}, a new sampling scheme that is designed to minimize the reconstruction error for inputs that lie within a known subspace while simultaneously bounding the maximum regret error for all other signals. The constrained sampling formulation leads to a convex combination of the subspace and the minimax regret samplings. It also yields an equivalent subspace sampling process with a modified input subspace. The constrained sampling is shown to be 1) (sub)optimal for signals in the input subspace, 2) robust for signals around the input subspace, 3) reasonably bounded for any signal in the entire space, and 4) flexible and easy to be implemented as combination of the subspace and regret samplings. We also presented a detailed theoretical analysis of reconstruction error of the proposed sampling. Additionally, we demonstrated the efficiency of constrained sampling through two illustrative examples. Our results suggest that the proposed sampling could potentially approach the optimum reconstruction (i.e., the orthogonal projection).  It would be intriguing to study the optimal selection of the parameter in the convex combination when more \emph{a priori} information about input signals become available.

\begin{appendices}

\section{Proof of Inequality~(\ref{eq-t1})}
\label{app-a}

As in the proof in~\cite[theorem $3$]{Eldar-2006},
we  represent any $x$ in $\D=\{x: \|x\|\leq L, c=S^*x\}$ as
\begin{eqnarray*}
x&=&P_\S x+ P_{\S^\bot}  x \\
&=&S(S^*S)^\dag c+v
\end{eqnarray*}
for some $v$ in $\G :=
 \{v\in \S^\bot : \|v\|^2 \leq  L^2-\|S(S^*S)^\dag c\|^2 \,\}$.

Let $a_c := WQc-P_\W S(S^*S)^\dag c$.
Then
\begin{eqnarray} \label{appendix-a}
\|R x\|^2 &= & \|P_\W x-WQS^*x\|^2\nn \\
&=& \|P_\W S(S^*S)^\dag c+P_\W v-WQc\|^2\nn\\
&=&  \|P_\W v-a_c\|^2\nn \\
&=& \|P_\W v\|^2- 2\mathrm{Re}\{\langle P_\W v, a_c\rangle\}+\|a_c\|^2 . \nn
\end{eqnarray}
Let
\[
v_1 := -\frac{\langle P_\W v, a_c\rangle}{\big|\langle P_\W v, a_c\rangle\big|} v.
\]
Clearly, $\|v_1\|=\|v\|$ and  $v_1\in \G$ if and only if  $v \in \G$.
Consequently
\begin{eqnarray*}
& & \max_{x\in \D} \|R x\|^2  \\
&=&\max_{v\in \G}\big\{\|P_\W v\|^2+ 2\big|\langle P_\W v, a_c\rangle\big|+\|a_c\|^2\big\} \\
&\geq &\|a_c\|^2+\max_{v\in \mathcal G}\|P_\W v\|^2 \\
& = &\|a_c\|^2+\max_{x\in \D}\|P_\W (x-P_\S x)\|^2 \\
&=&\|WQc-P_\W S(S^*S)^\dag c\|^2+
\max_{x\in \D}\|P_\W x-P_\W P_\S x\|^2 .
\end{eqnarray*}
On the other hand, since for any complex numbers $z_1$ and $z_2$,
\[
|z_1|^2+|z_2 |^2 \geq  \frac{1}{2}\big(\,|z_1|+|z_2|\,\big)^2,
\]
we get
\begin{eqnarray*}
  \max_{x\in \D}\|R x\|  &\geq &
\frac{1}{\sqrt 2} \bigl(\|WQc-P_\W S(S^*S)^\dag c\| \\
   & &  \hspace{5ex} +
\max_{x\in \mathcal D}\|P_\W x-P_\W P_\S x\| \bigr) .
\end{eqnarray*}
The proof is complete.
\hfill \IEEEQEDopen

\section{Proof of { Theorem~\ref{thm-con}}}
\label{app-b}

{
Let $c\in \mathcal{R}(S^*)$ be any given sample sequence. We first show that $\A \cap \D$ (in the objective function) contains only one element.
If $x\in \A $, then under the  direct-sum property  $\A\oplus\S^\perp=\H$, we have
\[
x = P_{\A \S^\perp} x = A(S^*A)^\dagger S^* x .
\]
On the other hand, if $ x\in \D$, then $S^* x =c$ according to the definition of $\D$ (\ref{sample-set}). Therefore,
\begin{equation}\label{eq-c9}
    x=A (S^*A)^\dag c .
\end{equation}
For the constraint, we denote the set of admissible correction filters that satisfy the regret constraint as}
\[
\mathcal{D}_Q := \{Q:\|P_\W S(S^*S)^\dagger c-WQc\|\leq \lam \beta (c)\}
\]
where $\beta (c)$ is given in~(\ref{upper-bound}),  $ \lam \in [0, 1] $.
The optimization problem in~(\ref{objective-final-subspace}) now becomes
\begin{equation}\label{eq-optimization}
\min_{Q\in \mathcal{D}_Q}\|A(S^*A)^\dagger c-WQc\|^2.
\end{equation}

Invoking orthogonal decomposition of $A(S^*A)^\dagger c-WQc$ onto $\W$ and $\W^\bot $
and using the triangular inequality, we have for any $Q\in \mathcal{D}_Q$, the objective function in (\ref{eq-optimization})  satisfy
\begin{eqnarray}\label{optimization-lower-subspace}
& & \|A(S^*A)^\dagger c-WQc\|^2\nn\\
& = &   \|P_\W A(S^*A)^\dagger c-WQc\|^2+\|P_{\W^\bot }A(S^*A)^\dagger c\|^2 \nn\\
& \ge &  \|P_{\W^\bot }A(S^*A)^\dagger c\|^2 +  \Big|\|P_\W S(S^*S)^\dagger c-WQc\| \nn\\
&& \hspace{4ex} -\|P_\W S(S^*S)^\dagger c -P_{\W }A(S^*A)^\dagger c \|\Big|^2 \nn\\
&=&  \Big|\|P_\W S(S^*S)^\dagger c-WQc\|-\beta (c)\Big|^2  +\|P_{\W^\bot }A(S^*A)^\dagger c\|^2\nn\\
& \geq &(1-\lam)^2 {\beta^2(c)}+\|P_{\W^\bot }A(S^*A)^\dagger c\|^2.
\end{eqnarray}
Substituting
\[
Q=\lam Q_\sub+(1-\lam)Q_\reg
\]
into~(\ref{eq-optimization}), we see that the lower bound in~(\ref{optimization-lower-subspace}) is reached.
That completes the proof. \hfill \IEEEQEDopen

\section{Proof of~Proposition~\ref{prop-x2}}
\label{app-p2}
Since $P_{\A \S^\bot }$ and $P_\S$ have the same nullspace $\S^\bot$,
applying Proposition~\ref{prop-1} on $B$ in (\ref{eq-t4}) concludes $B$ is also an projection.

It remains to be shown that  $\NS (B)=\S^\bot$.
It suffices if we show that $ B x=0$ if and only if $P_\S x=0$,
which can be proved by an alternative expression of $B$ (in terms of $P_\S$ and $P_{\S^\bot \A}$):
\begin{eqnarray}\label{BB}
  B  &=& \lam  P_{\A\S^\bot } + (1-\lam)  P_\S\nn \\
   &=&  \lam  P_{\A\S^\bot } + (1-\lam)  P_\S P_{\A\S^\bot }\nn   \\
   &=&  [\lam  I+ (1-\lam)P_\S]  P_{\A\S^\perp}\nn   \\
   &=&  [P_\S+\lam (I- P_\S)]  P_{\A\S^\bot }\nn   \\
   &=& [P_\S+\lam   P_{\S^\perp}]  P_{\A\S^\bot }\nn   \\
   &=& P_{\S} + \lam   P_{\S^\bot} P_{\A\S^\bot }
\end{eqnarray}
where the second step is from~(\ref{projection1}), the second to the
last step is due to~(\ref{eq-x1}),  and the last step is from~(\ref{projection1}).
For any $x\in \H$, since $P_\S x $ and $P_{\S^\bot} P_{\A\S^\bot } x$ are perpendicular to each other, the statement then follows immediately. The proof is complete.  \hfill \IEEEQEDopen

\section{Proof of Bounds of $\cos (\B,\S) $ in~(\ref{inequality-cos})}
\label{app-a1}

Since $\NS(B)=\S^\bot$, we have from (\ref{angle-c}) that
\begin{equation}\label{eq-a0}
\cos^2 \big(\B,\S \big)= \inf_{x \not\in \S^\bot} f(x)
\end{equation}
where
\begin{equation}\label{eq-ap1}
  f(x) := \frac{\|P_\S  B  x\|^2}{\|B x\|^2} .
\end{equation}
Since $  B=P_\S+\lambda P_{\S^\perp} P_{\A \S^\perp}$  (see~(\ref{BB})), thus
\begin{eqnarray}
f(x)  &=&
\frac{ { \|P_\S  ( P_\S+\lambda P_{\S^\perp} P_{\A \S^\perp})x \|^2}}
           {\|P_\S x+\lambda P_{\S^\perp} P_{\A \S^\perp} x \|^2}\nn \\
&=&  \frac{\| P_\S x\|^2}{\|P_\S x\|^2 +\lam^2  \|P_{\S^\perp} P_{\A \S^\perp} x  \|^2 }  \nn \\
&=&   \frac{1}{1+\lam^2 \frac{ \|P_{\S^\perp} P_{\A \S^\perp} x \|^2}{\|P_\S x\|^2}}
\label{eq-a2}
\end{eqnarray}
where the second step for the denominator is due to the
orthogonality of $P_{\S^\perp} P_{\A \S^\perp} x$ to $P_\S x$.
From~(\ref{eq-x12}), it holds
\begin{equation}\label{bound-two-orthogonal}
\cos  (\A, \S^\perp) \|P_{\A \S^\perp} x\| \leq {\|P_{\S^\perp} P_{\A \S^\perp} x\|} \leq \sin ( \A,\S )\|P_{\A \S^\perp} x\|.
\end{equation}
Then, from~(\ref{consistent-bound}), it follows that
$P_{\A \S^\perp} x$ satisfies
\begin{equation}\label{bound-oblique-orthogonal}
\frac{\|P_\S x\|}{\sin (\A, \S^\perp  )}
\leq \|P_{\A \S^\perp} x\|
\leq \frac{\|P_\S x\|}{\cos\bigl(\A,\S\bigr)}.
\end{equation}
Combining~(\ref{bound-two-orthogonal}) and~(\ref{bound-oblique-orthogonal}) yields
\begin{equation}\label{eq-a5}
\frac{\cos (\A, \S^\perp  )}{\sin (\A, \S^\perp )} \|P_\S x\|
\leq \|P_{\S^\perp} P_{\A \S^\perp} x\|
\leq \frac{\sin(\A,\S)}{\cos\bigl(\A,\S\bigr)} \|P_\S x\|.
\end{equation}
As a result, we have from (\ref{eq-a2}) that
\begin{equation}\label{eq-a7}
\frac{1}{1+\lam^2\frac{\sin^2(\A,\S)}{\cos^2(\A,\S)}}
\leq f(x)
\leq \frac{1}{1+\lam^2\frac{ \cos^2 (\A, \S^\perp )}{\sin^2 (\A, \S^\perp  )}}.
\end{equation}
Then (\ref{inequality-cos}) follows immediately from (\ref{eq-a0}) and (\ref{eq-a7}).
\hfill \IEEEQEDopen

\section{Proof Bounds of $\sin (\B^\bot,\S) $ in~(\ref{inequality-sin}) }
\label{app-a2}

Since $\NS(P_{\B^\bot  \S})=\S$, we have from (\ref{angle-s}) that
\begin{equation}\label{eq-a1}
  \sin^2\big (\B^\bot ,\S\big) =  \sup_{x\not \in \S} g(x)
\end{equation}
where
\begin{equation}\label{eq-x97}
g(x) :=
    \frac{\|P_{\S^\bot } P_{\B^\bot  \S}x\|^2}{\|P_{\B^\bot  \S}x\|^2}  .
\end{equation}
According to~\cite{Christensen-2004}, the adjoint operator of any projection $P_{\V_1 \V2}$ is also a projection
and further we have
\begin{equation}\label{eq-x4}
  P_{\V_1 \V_2}^* = P_{\V_2^\perp \V_1^\perp}.
\end{equation}
Hence,
\begin{eqnarray*}
P_{\B^\bot  \S}&=&I-P_{\S\B^\bot }\nn\\
&=&I-B^* \nn\\
&=&I-\bigl(\lam P_{\A\S^\bot } + (1-\lambda) P_\S)^* \\
&=&I-\bigl(\lam P_{\S \A^\bot } + (1-\lambda) P_\S \bigr) \\
& = & \lambda P_{\A^\perp \S} + (1-\lambda) P_{\S^\perp}\\
& = & \lambda P_{\A^\perp \S} + (1-\lambda) P_{\S^\perp} P_{\A^\perp \S}\\
& = & [\lambda I + (1-\lambda) P_{\S^\perp}] P_{\A^\perp \S}\\
& = & [ P_{\S^\perp} + \lambda P_\S ] P_{\A^\perp \S}\\
&=& P_{\S^\perp} +  \lambda P_\S P_{\A^\perp \S}.
\end{eqnarray*}
Note that $g(x)$ in (\ref{eq-x97}) has the same form as $f(x)$ in (\ref{eq-ap1}),  except that all the subspaces involved are replaced by their respective orthogonal complements.
Using~(\ref{eq-a7}) and noting $(\A^\bot,\S^\bot)=(\S,\A) =(\S, \A)$. We finally obtain
\begin{equation*}
\frac{1}{1+\lam^2\frac{\sin^2(\A, \S )}{\cos^2(\A, \S)}}
\leq g(x)
\leq \frac{1}{1+\lam^2\frac{\cos^2(\A^\bot,\S)}{\sin^2(\A^\perp ,\S)}} .
\end{equation*}
Then, inequality~(\ref{inequality-sin}) follows immediately.
 \hfill \IEEEQEDopen
\end{appendices}



\end{document}